\documentstyle[aps,prl,multicol,epsf,floats,epsfig,amsmath,amssymb]{revtex}

\def\limo{\mathrel{\mathop{\longrightarrow}\limits_{n\to \infty}}}

\begin{document}
  
\draft 

\title
{A consistent Lie algebraic representation of \\ quantum phase and number operators}
\author {M. Rasetti$^{\#\, ,\circledast}$}
\address{$^{\#}$Dipartimento di Fisica, Politecnico di Torino, I-10129 Torino, Italy \\ 
$^{\circledast}$Istituto Nazionale di Fisica della Materia, Unit\`a Politecnico di 
Torino, I-10129 Torino, Italy} 
\maketitle

\begin{abstract}
A consistent realization of the quantum operators corresponding to the canonically 
conjugate phase and number variables is proposed, resorting to the $\kappa =
\frac{1}{2}$ positive discrete series of the irreducible unitary representation of 
the Lie algebra $su(1,1)$ of the double covering group of $SO^{\uparrow}(1,2)$.  
\end{abstract}


\begin{multicols}{2}[]
\narrowtext
 
The problem -- quite relevant both for quantum optics and the physics of mesoscopic 
systems (Josephson junctions) -- of how to represent the canonically conjugate phase 
${\hat{\phi}}$ and number ${\hat{n}}$ operators ($[ {\hat{\phi}} , {\hat{n}} ] =i$) 
as self-adjoint operators in the Hilbert ({\sl e.g.} Fock) space ${\mathfrak{F}}$ of 
states of the corresponding physical system remains essentially unsolved. The several 
attempts made over the years \cite{Di} --  
\cite{WeVoOp}, have clarified a number of interesting features and have proposed 
practical approaches to deal with various obstructions, but none has fully overcome 
the difficulties, inherent essentially to the inconsistency related with the definition 
of the phase in the state where the number eigenvalue is zero (a difficulty which 
has a classical counterpart \cite{Go}). 

The correspondence principle of quantum mechanics appears to suggest that the 
customary annihilation and creation operators $a$, $a^{\dagger}$ (related to the complex 
classical amplitude of harmonic oscillations) adopted {\sl e.g.}, in the description of 
the harmonic oscillator, which together with the number ${\hat{n}}$ and identity operator 
${\mathbb{I}}$, generate the Weyl-Heisenberg algebra $h(1)$ $\bigl ( \displaystyle{\left 
[\, a , a^{\dagger} \right ] = {\mathbb{I}}}$, $\displaystyle{\left [\, a , {\hat{n}}\, 
\right ] = a}$, $\displaystyle{\left [ a^{\dagger} , {\hat{n}}\, \right ] = - a^{\dagger}}$, 
$\displaystyle{\left [ \bullet , {\mathbb{I}} \, \right ] = 0} \bigr )$, may be represented 
in polar form (phase-modulus decomposition) as 
\begin{eqnarray}
a \doteq {\hat{e}} \, \sqrt{{\hat{n}}} \; , \label{mod-phas} 
\end{eqnarray} 
${\hat{e}}$ defining the quantum phase operator. However, it is not possible to 
derive from the above ansatz a fully satisfactory phase observable: indeed, the solution 
of (\ref{mod-phas}) $\displaystyle{{\hat{e}} = \sum_{n=0}^{\infty} |n\rangle
\langle n+1|}$, where ${\hat{n}}\, |n\rangle = n\, |n\rangle \, n\geq 0$, 
turns out not to be unitary; 
\begin{eqnarray}
{\hat{e}}\, {\hat{e}}^{\dagger} = {\mathbb{I}} \; , \; {\hat{e}}^{\dagger} {\hat{e}} = 
{\mathbb{I}} - |0\rangle\langle 0| \; . \label{commph} 
\end{eqnarray}

The approach to the problem most adopted in the applications is that due to Pegg and 
Barnett \cite{PeBa}, who propose to resort to a finite dimensional \cite{others} Hilbert space of 
of states ${\cal H}_N \doteq {\rm span} \{ |n\rangle \, |\, n=0,\dots ,N-1 \}$ ($N = {\rm 
dim}\, ({\cal H}_N)$) and to define in it a self adjoint phase operator ${\hat{\phi}}_N$ 
by constructing first the set of $''$phase states$\, ''$
\begin{eqnarray}
|\phi_{\ell}\rangle \doteq N^{-\frac{1}{2}}\, \sum_{n=0}^{N-1} \exp \left ( i\, n\, 
\phi_{\ell} \right ) \, |n\rangle \; , \; \phi_{\ell} \doteq \phi_0 + 2\pi \frac{\ell}{N} 
\; , \label{ph-states}   
\end{eqnarray}
for $\ell \in {\mathbb{Z}}_N$, and setting then $\displaystyle{{\hat{\phi}}_N \doteq 
\sum_{\ell =0}^{N-1} \phi_{\ell} \, |\phi_{\ell}\rangle\langle \phi_{\ell} |}$. 
In the occupation number basis ${\hat{\phi}}_N$ has matrix elements 
\begin{eqnarray}
\langle \ell \, |{\hat{\phi}}_N |\, \ell \, \rangle &=& \phi_0 + \pi \, (N-1) N^{-1} \quad 
{\rm and}\, , \; {\rm for}\; m\neq \ell \; ,\nonumber \\ 
\langle m \, |{\hat{\phi}}_N |\,\ell \,\rangle &=& \frac{2\pi^2}{N} \, \left | \sin \left ( 
 \frac{\pi}{N} \, (m-\ell )\right ) \right |^{-1} \, {\rm e}^{i(m - \ell )\left ( \phi_0 + 
\frac{1}{N}\, \pi \right )} \; . \nonumber  
\end{eqnarray}
The problem here is that the spectral resolution of the discrete operator 
${\hat{\phi}}_N$ does not provide \cite{Pa} a measure converging either to a 
projection valued measure 
nor to a probability operator measure 
in the limit $N\to \infty$ ({\sl i.e.} in the full Hilbert-Fock space ${\mathfrak{F}}$). 

The most promising approaches to circumvent such difficulties came from the identification 
of the algebraic structure underlying the problem. Ref. \cite{Ell} resorts first to the polar 
decomposition of step operators in $u(2)$ $\equiv u(1)\otimes su(2)$ and then to its contraction 
limit. Being $su(2)$ compact, the closure property in this way is lost. At this point the 
idea of using $su(1,1)$, non-compact counterpart of $su(2)$, was introduced \cite{BaPe}, \cite{Ban},  
based on relative number variables, well defined ({\sl e.g.} in the frame of thermo-field 
dynamics) in a two-mode representation of $su(1,1)$. The crucial notion in these approaches was 
just the realization of the role of such algebra. This notion was recently revived. Observing that 
in a canonical quantization scheme the self-adjoint Lie algebra generators $K_1$, $K_2$, $K_3$ of the 
group $SO^{\uparrow}(1,2)$ correspond to the classical polar coordinates variables in 
${\mathbb{R}}^2$, ${\cal R}_x \doteq r \cos \phi$, ${\cal R}_y \doteq r \sin \phi$ and 
${\cal R}_z \doteq r$, respectively, whose Poisson brackets $\bigl ( \{ {\cal R}_x , {\cal 
R}_y \}_{PB} = {\cal R}_z$ , $\{ {\cal R}_z , {\cal R}_x \}_{PB} = - {\cal R}_y$ , $\{ {\cal R}_z , 
{\cal R}_y \}_{PB} = {\cal R}_x \bigr )$ satisfy the same algebra, Kastrup \cite{Ka} proposed a 
group theoretical approach to the problem resorting to the irreducible unitary representations of the 
positive series.  
 
The generators $K_1$, $K_2$, $K_3$ have commutation relations $\displaystyle{\left 
[ K_1 , K_2 \right ] = -i K_3}$, $\displaystyle{\left [ K_2 , K_3 \right ] = i K_1}$, 
$\displaystyle{\left [ K_3 , K_1 \right ] = i K_2}$, or, introducing the skew raising and 
lowering operators $K_{\pm} \doteq K_1 \pm i K_2$ acting as ladder operators on the 
eigenvectors of the Cartan operator $K_3$, generator of the compact subgroup $SO(2)$ of 
$SO^{\uparrow}(1,2)$, $\displaystyle{\left  [ K_+ , K_- \right ] = -2 K_3 \; , \; \left [ K_3 , 
K_{\pm} \right ] = \pm K_{\pm}}$.

The positive discrete series representation of the algebra $su(1,1)$ of the double 
covering of $SO^{\uparrow}(1,2)$ is characterized \cite{Pe} by the existence of an 
highest weight vector $|\kappa ,\Omega \rangle$ (invariant, in representation space, under the 
action of the maximal compact sub-group ${\cal K}\sim U(1)$ of $SU(1,1)$ \cite{su(11)}), annihilated 
by the lowering operator $K_-$. This means that  
$K_- |\kappa ,\Omega \rangle$ is \underline{not} 
a ket in the representation Hilbert space. Upon identifying $|\kappa ,\Omega \rangle$ with the 
eigenvector $|\kappa ,0 \rangle$ of $K_3$ the commutation relations give $K_- |\kappa 
,0 \rangle \equiv 0$ (see below). The real number $\kappa$, characterizing the representation, 
assumes values $\kappa = \frac{1}{2}, 1, \frac{3}{2}, 2, \dots $, and is such that the Casimir 
operator ${\cal C}_2 = K_1^2 + K_2^2 - K_3^2$ has in that representation eigenvalue $\kappa (1-\kappa )$. 
The positive discrete series irreducible unitary representation ${\cal D}^{(+)}_{\kappa}$, 
with $K_-^{\dagger} \equiv K_+$, is spanned by the complete orthonormal set $\{ |\kappa ,n 
\rangle \, |\, n \in {\mathbb{N}} \}$ of eigenstates of $K_3$, $K_3 | \kappa ,n\rangle = 
\left ( n + \kappa \right )\, | \kappa ,n\rangle$.  
The ladder operators $K_+$, $K_-$ satisfy the equations 
\begin{eqnarray}
K_+ |\kappa ,n\rangle &=& \omega_n \left ( n+1 \right )\, |\kappa 
,n+1\rangle \; , \nonumber \\ 
K_- |\kappa ,n\rangle &=& \omega_{n-1}^{-1} n \, 
|\kappa ,n-1\rangle \; , \nonumber 
\end{eqnarray} 
where $\omega_n = {\rm e}^{i \theta_n}$ is a phase. With no loss of generality $\theta_n$ 
can be assumed independent on $n$ ($\omega_n = {\rm e}^{i \theta}\doteq \omega \, ,\, \forall 
n\in {\mathbb{N}}$). The equations above imply 
\begin{eqnarray}
|\kappa , n\rangle = \omega^{-n} \frac{1}{n!} \, K_+^n \, 
|\kappa , 0\rangle \; . \label{power} 
\end{eqnarray}
${\cal D}^{(+)}_{\frac{1}{2}}$ has an interesting realization in $h(1)$: the basis vector 
$|\frac{1}{2} ,n\rangle$ can be identified with the eigenvector $|n\rangle$ of ${\hat{n}}$ 
in ${\mathfrak{F}}$, and over ${\mathfrak{F}}$ 
\begin{eqnarray}
K_+ \equiv {\hat{n}}^{\frac{1}{2}}
\, a^{\dagger} \; , \; K_- \equiv a \, {\hat{n}}^{\frac{1}{2}}
\; , \; K_3 \equiv {\hat{n}} + {\scriptstyle{\frac{1}{2}}} 
\; . \nonumber 
\end{eqnarray}

There is a connection between this representation of $su(1,1)$ and the 
uniform expansion to which (\ref{ph-states}) reduces for $\ell =0$, which is at the 
basis of the lack of covergence mentioned above: a uniform superposition of occupation 
number eigenstates in ${\mathfrak{F}}$ can be straightforwardly constructed in ${\cal 
D}_{\frac{1}{2}}^{(+)}$, in view of (\ref{power}), as \cite{RaTaZe} $\displaystyle{{\rm 
e}^{\omega K_+} |{\scriptstyle{\frac{1}{2}}} , 0\rangle = \sum_{n=0}^{\infty} 
|{\scriptstyle{\frac{1}{2}}} , n\rangle}$, a state which is manifestly \underline{not} 
normalizable. 

For analogy with the classical case $\bigl ($ $\cos \phi = r^{-1} {\cal R}_x$, $\sin \phi 
= r^{-1} {\cal R}_y$ $\bigr )$ in \cite{Ka} the definition is suggested:   
\begin{eqnarray}
{\widehat{\cos \phi}} \equiv {\scriptstyle{\frac{1}{2}}} \left \{ K_3^{-1} , K_1 \right \} 
\; , \; {\widehat{\sin \phi}} \equiv {\scriptstyle{\frac{1}{2}}} \left \{ K_3^{-1} , K_2 \right \} 
\; . \label{cossin} 
\end{eqnarray} 
In the discrete positive series irreducible unitary representation of $su(1,1)$, this 
assumption is analytic for generic $\kappa$, because $K_3^{-1}$ is well defined in 
view of the eigenvalue equation of $K_3$, but it has the drawback that ${\widehat{\cos 
\phi}}^2 + {\widehat{\sin \phi}}^2$ and $\bigl [ \, {\widehat{\cos \phi}} , 
{\widehat{\sin \phi}} \, \bigr ]$, even though both diagonal, are not equal -- 
respectively -- to ${\mathbb{I}}$ and $0$, as they should be.  Indeed, they do satisfy 
such conditions only for large $n$: 
\begin{eqnarray}
 && \left ( {\widehat{\cos \phi}}^2 + {\widehat{\sin \phi}}^2 \right )  
|\kappa ,n\rangle = \frac{1}{8} \left [ \left ( f_{n+1}^{(\kappa )} 
\right )^2 + \left ( f_n^{(\kappa )} \right )^2 \right ] |\kappa ,n\rangle \, , 
\nonumber \\ 
 && \bigl [ \, {\widehat{\cos \phi}} , {\widehat{\sin \phi}} \, \bigr ] \, |\kappa 
,n\rangle = \frac{1}{8i} \left [ \left ( f_{n+1}^{(\kappa )} \right )^2 
- \left ( f_n^{(\kappa )} \right )^2 \right ] |\kappa ,n\rangle \, , \label{sqco} 
\end{eqnarray}  
where 
\begin{eqnarray}
f_n^{(\kappa )} \doteq \left [ n \, (n+2\kappa -1) \right ]^{\frac{1}{2}}\, \left ( 
\frac{1}{n+\kappa } + \frac{1}{n+\kappa -1} \right ) \limo 2  \; . \nonumber 
\end{eqnarray}
In this note we suggest that the above scheme can be generalized so as to exhibit 
all the required features for effective phase-angle quantum variables. 

The basic step is the generalization of eqs. (\ref{cossin}): 
\begin{eqnarray}
{\mathfrak{c}} \equiv {\widehat{\cos \phi}} \doteq \left \{ {\cal F}(K_3) , K_1 \right \} 
\, , \, {\mathfrak{s}} \equiv {\widehat{\sin \phi}} \doteq \left \{ {\cal F}(K_3) , K_2 
\right \} \, , \label{cosi}   
\end{eqnarray}
where ${\cal F}$ is a function -- whose existence will be proved in the sequel 
-- meromorphic in the ${\cal D}^{(+)}_{\frac{1}{2}}$-representation,  
of the form ${\cal F}(K_3 ) = {\hat{n}}^{-1}\bigl ( {\mathbb{I}} - |0\rangle\langle 0| 
\bigr ) + \Phi ({\hat{n}})$, 
where $\Phi ({\hat{n}})$, 
whose eigenvalues in ${\mathfrak{F}}$ will be written in terms of digamma functions, is 
such that $\Phi ({\hat{n}}) \, |0\rangle =0$. Such choice guarantees that the commutation 
relations $\displaystyle{\left [ {\hat{n}} , {\widehat{\cos \phi}} \right ] = -i \, 
{\widehat{\sin \phi}}}$, $\displaystyle{\left [ {\hat{n}} , {\widehat{\sin \phi}} \right ] = i \, 
{\widehat{\cos \phi}}}$, are satisfied $\bigl ( {\hat{n}} = K_3 - \frac{1}{2} \bigr )$ and 
${\cal F}$ may be selected so as to ensure that that the two other conditions are satisfied 
almost everywhere ({\sl i.e.} in ${\mathfrak{F}} \setminus |0\rangle$). The construction ${\cal F}$ 
in ${\cal D}^{(+)}_{\frac{1}{2}}$ in such a way that : 
\begin{eqnarray} 
{\rm i)}\quad {\widehat{\cos \phi}}^2 + {\widehat{\sin \phi}}^2 = {\mathbb{I}} \quad , 
\quad {\rm ii)}\quad \bigl [ \, {\widehat{\cos \phi}} , {\widehat{\sin \phi}} \, \bigr ] 
= 0 \; , \label{cond}  
\end{eqnarray} 
proceeds as follows. 
One observes first $\bigl ( |n\rangle\equiv |\frac{1}{2} ,n\rangle \bigr )$ that 
\begin{eqnarray}
{\mathfrak{c}} \, |n\rangle &=& \omega \, {\cal L}_{n+1} \, |n+1\rangle + \omega^{-1}\,  
{\cal L}_n \, |n-1\rangle \; , \nonumber \\ 
{\mathfrak{s}} \, |n\rangle &=& -i\, \omega \, {\cal L}_{n+1} \, |n+1\rangle + i\, 
\omega^{-1} \, {\cal L}_n \, |n-1\rangle \; , \nonumber 
\end{eqnarray}
where ${\cal L}_n \doteq \frac{1}{2}\, n\, \bigl ( {\cal F} \left (n - 
\frac{1}{2} \right ) + {\cal F} \left (n + {\scriptstyle{\frac{1}{2}}} 
\right )\bigr )$.
Equivalently, 
\begin{eqnarray}
{\mathfrak{e}}_{\pm} \doteq \left ( {\mathfrak{c}} \pm i\, {\mathfrak{s}} \right ) 
\equiv {\widehat{{\rm e}^{\pm i\phi}}} \; , \; {\mathfrak{e}}_{\pm} = \left \{ {\cal F}(K_3 ) ,
K_{\pm} \right \} \; , \nonumber 
\end{eqnarray}
satisfy 
\begin{eqnarray}
{\mathfrak{e}}_+ \, |n\rangle = 2 \omega \, {\cal L}_{n+1}\, |n+1\rangle \; , 
\; {\mathfrak{e}}_- \, |n\rangle = 2 \omega^{-1} \, {\cal L}_n \, 
|n-1\rangle \, . \nonumber 
\end{eqnarray} 
From these, for 
\begin{eqnarray}
{\cal O} \doteq \frac{1}{4} \xi \, \left [ (1+\eta )\,  
{\mathfrak{c}} + i (1-\eta ) \, {\mathfrak{s}} \right ]\, \left [ (1+\xi \eta )\, 
{\mathfrak{c}} -i (1-\xi \eta ) \, {\mathfrak{s}} \right ] \; , \nonumber 
\end{eqnarray} 
where $\xi ,\eta \in \{ \pm 1 \}$, $\bigl ( {\cal O}_{1,1} 
\equiv {\mathfrak{c}}^2$, ${\cal O}_{1,-1} \equiv {\mathfrak{s}}^2$, ${\cal O}_{-1,1} 
\equiv -i {\mathfrak{s}} {\mathfrak{c}}$, ${\cal O}_{-1,-1} \equiv i {\mathfrak{c}} 
{\mathfrak{s}} \bigr )$, one finds    
\begin{eqnarray}
{\cal O}_{\xi ,\eta}&& \!\!\!\! |n \rangle = \eta \left [\, {\cal L}_n^2 + \xi {\cal L}_{n+1}^2
\right ] \, |n\rangle \nonumber \\ &+& \eta \bigl ( \omega^2 \, {\cal L}_{n+1} {\cal L}_{n+2} 
\, |n+2\rangle + \omega^{-2}\, {\cal L}_{n-1} {\cal L}_n \, |n-2\rangle \bigr ) 
\; , \nonumber  
\end{eqnarray}
whence $\displaystyle{\left ({\mathfrak{c}}^2 + {\mathfrak{s}}^2 
\right )\, |n\rangle = 2\, \left [\, {\cal L}_n^2 + {\cal L}_{n+1}^2\right ] \, |n\rangle}$, 
and $\displaystyle{\left [\, {\mathfrak{c}} , {\mathfrak{s}} \, \right ]\, |n\rangle }$ 
$\displaystyle{= 2i\, \left [\, {\cal L}_{n+1}^2 - {\cal L}_n^2\right ] \, |n\rangle}$. 
The pursued conditions (\ref{cond}) amount then to requiring that ${\cal L}_{n+1} = {\cal L}_n 
= \frac{1}{2}$, for all $n\geq 1$, namely that the following system of 
recursion equations is satisfied by $F(n) \doteq {\cal F}\left ( n+ \frac{1}{2}\right )$: 
\begin{eqnarray}
~~&~& (n+1) F(n+1) - n F(n-1) + F(n) = 0 \; , \label{recur11} \\  
~~&~& F(n) = \frac{1}{n} - F(n-1) \; . \label{recur22} 
\end{eqnarray}
Eqs. (\ref{recur22}) and (\ref{recur11}) are mutually consistent; thus we only need to consider, 
{\sl e.g.}, (\ref{recur11}). Eq. (\ref{recur22}) will simply be used to properly continue the result 
to $n=0$. For $n\geq 1$, recursion equation (\ref{recur11}) can be dealt with by the generating function 
method. One defines first   
\begin{eqnarray}
f(z) \doteq \sum_{n=1}^{\infty} F(n) z^n \; , \label{f}  
\end{eqnarray} 
where $z$ is an undeterminate. Multiplying eq. (\ref{recur11}) by $z^n$ and summing 
for $n$ ranging from 1 to $\infty$, in view of definition (\ref{f}) one obtains for 
$f$ the differential equation 
\begin{eqnarray}
\left ( 1-z^2 \right )\, \frac{df}{dz} + \left ( 1-z \right )\, f(z) - F(1) = 0 
\; . \label{eqdiff}  
\end{eqnarray}
In view of (\ref{recur22}) we select $F(1)=1$, so that, as only one integration constant 
is required, we may set $F(0)\equiv f(0) = 0$. The solution of (\ref{eqdiff}) is then  
\begin{eqnarray}
f(z) = - (1+z)^{-1} \, \ln (1-z) \; . \label{sol} 
\end{eqnarray}  
Expanding (\ref{sol}) as a power series in $z$ one eventually finds 
\begin{eqnarray}
F(n) = (-)^n \sum_{\ell =1}^n \frac{(-)^{\ell}}{\ell} \; , \; \forall\, n\geq 1 \; . 
\label{Fn}  
\end{eqnarray}
The sum in (\ref{Fn}) can be split into even and odd values of $\ell$, and has 
different expressions depending on whether $n$ is even or odd: 
\begin{eqnarray}  
F(2m) &=& {\scriptstyle{\frac{1}{2}}} \, \sum_{r=1}^m \frac{1}{r} - \sum_{r=o}^{m-1} 
\frac{1}{2r+1} \; , \; m\geq 1\; , \nonumber \\  
F(2m+1) &=& -{\scriptstyle{\frac{1}{2}}} \, \sum_{r=1}^m \frac{1}{r} + \sum_{r=0}^m 
\frac{1}{2r+1} \; , \; m\geq 0 \; . \nonumber 
\end{eqnarray}
Recalling now that \cite{AbSt} $\displaystyle{\sum_{\ell =1}^k \frac{1}{\ell} = \gamma 
+ \psi (k+1)}$, and $\displaystyle{\sum_{\ell =0}^{k-1} \frac{1}{2\ell +1} = 
{\scriptstyle{\frac{1}{2}}} \left (\gamma + 2 \ln 2 + \psi \left (k+{\scriptstyle{\frac{1}{2}}}
\right ) \right )}$, where $\gamma$ is Euler's constant and $\psi (z)$ the Digamma function, and 
that $\displaystyle{{\scriptstyle{\frac{1}{2}}} \psi \left ( k+{\scriptstyle{\frac{1}{2}}} \right ) 
= \psi (2k) - {\scriptstyle{\frac{1}{2}}} \psi (k) - \ln 2}$, one obtains -- including as well the 
condition $F(0)=0$, and denoting by $[\![ x ]\!]$ the largest integer $\leq x$ -- 
\begin{eqnarray}
F(n) &=& \left ( 1 - \delta_{n,0} \right )\, \frac{1}{n} + \Phi (n) \; , \nonumber \\ 
\Phi (n) &=& (-)^n \left [ \psi \left ( \left [\!\left [ {\scriptstyle{\frac{1}{2}}} 
(n+1) \right ]\!\right ] \right ) - \psi (n) \right ]\; . \label{Fnf}    
\end{eqnarray}
 
We may then conclude that the function entering definition (\ref{cosi}) ${\cal F}(K_3) 
\equiv F({\hat{n}})$, does exist and is well defined as operator \cite{note} over 
the Fock space ${\mathfrak{F}}$ for $n\geq 1$. With this choice, indeed,  
\begin{eqnarray}
{\mathfrak{e}}_+ \, |n\rangle = \omega \, |n+1\rangle \; ,\; n\geq 0 \; ; \; 
{\mathfrak{e}}_- \, |n\rangle = \omega^{-1} \, |n-1\rangle \; , \; n\geq 1 \; .  
\nonumber 
\end{eqnarray}
Still what happens in state $|0\rangle$ has some subtlety which requires further 
discussion. Explicit calculation using the scheme discussed above shows that while the 
conditions (\ref{cond}) hold for all $|n\rangle$ with $n\geq 1$, one has $\bigl ( 
{\widehat{\cos \phi}}^2 + {\widehat{\sin \phi}}^2 \bigr )\, |0\rangle = \frac{1}{2} \, 
|0\rangle$, $\bigl [ \, {\widehat{\cos \phi}} , {\widehat{\sin \phi}} \, \bigr ]\, 
|0\rangle = - \frac{1}{2} i \, |0\rangle$ and ${\mathfrak{e}}_- \, |0\rangle = 0$. This 
is the $''$fossil remain$\, ''$ in our representation of the pathology encountered in all 
other representations and the price paid for making independent on $n$ the eigenvalues in 
(\ref{sqco}). The action of ${\mathfrak{e}}_-$ is badly defined in state 
$|0\rangle$: one finds ${\mathfrak{e}}_+ {\mathfrak{e}}_- = {\mathbb{I}} - |0\rangle 
\langle 0|$, ${\mathfrak{e}}_- {\mathfrak{e}}_+ = {\mathbb{I}}$, namely $\bigl [ 
{\mathfrak{e}}_- , {\mathfrak{e}}_+ \bigr ] = |0\rangle\langle 0|$, the same disease as 
in (\ref{commph}), even though within a much more robust structure, where sine (${\mathfrak{s}}$) 
and cosine (${\mathfrak{c}}$) operators satisfy the desired conditions ${\mathfrak{c}}^2 + 
{\mathfrak{s}}^2 = {\mathbb{I}}$, $\bigl [ \, {\mathfrak{c}} , {\mathfrak{s}} \, \bigr ] = 0$ 
in ${\mathfrak{F}}\setminus |0\rangle$. The message is clear: in state $|0\rangle$ operator 
${\hat{n}}$ is sharp; $\Delta_0 ({\hat{n}}) = \bigl [ \langle 0|{\hat{n}}^2|0 \rangle - 
\langle 0|\,{\hat{n}}\, |0 \rangle^2 \bigr ]^{\frac{1}{2}} = 0$, then $\Delta_0 
({\hat{\phi}})$ can be arbitrarily large. We find $\Delta_0 ({\mathfrak{c}}) = 
\Delta_0 ({\mathfrak{s}}) = \frac{1}{\sqrt{2}}$, which corresponds to the minimum uncertainty 
condition $\Delta_0 ({\mathfrak{c}}) \, \Delta_0 ({\mathfrak{s}}) = \frac{1}{2}$ for the 
commutation relation $\bigl [ \, {\mathfrak{c}} , {\mathfrak{s}} \, \bigr ]\, 
|0\rangle = - \frac{1}{2} i \, |0\rangle$.        

The new feature here, however, is that (\ref{Fnf}) can be analytically continued to $n=-1$, where, 
in view of the identity \cite{AbSt} $\psi (1-z) = \psi (z) + \pi \cot (\pi z)$, one finds $F(-1)=0$. 
Notice that this is not consistent with (\ref{recur22}), which however obviously does not hold for 
$n=0$. For $n=0$ a more plausible solution to (\ref{recur22}) would instead demand  
\begin{eqnarray}
\lim_{n \to 0} n F(n-1) = 1 \; . \label{lim} 
\end{eqnarray}  
This leads to an intriguing possible way out of our difficulty, that consists in 
replacing $\Phi (n)$ in (\ref{Fnf}),   
\begin{eqnarray}
\Phi (n) \Rightarrow && (-)^n \bigl [ \psi \left ( \left [\!\left 
[ {\scriptstyle{\frac{1}{2}}} (n+1) \right ]\!\right ] \right ) - 
\psi (n) - 2\, \delta_{n,-1} \psi (n+1) \nonumber \\ && + \, \delta_{n,-2} \left [ 
\left ( \gamma - {\scriptstyle{\frac{1}{2}}} \right ) (n+2) -1 \right ] 
\psi (n+2) \bigr ] \; . \label{Fi}     
\end{eqnarray}
Indeed, with this new self-adjoint operator $\Phi({\hat{n}})$, $F({\hat{n}})$, as 
given by the first of eqs. (\ref{Fnf}), leads to ${\mathfrak{e}}_+ {\mathfrak{e}}_- 
= {\mathbb{I}}$, ${\mathfrak{e}}_+ = {\mathfrak{e}}_-^{\dagger}$ and $\bigl [ 
{\mathfrak{e}}_- , {\mathfrak{e}}_+ \bigr ] =0$ also in state $|0\rangle$. 

There is a price to be paid, of course. What the above contruction requires is a non trivial 
mathematical structure, implying the extension of the space of states ${\mathfrak{F}}$ to include 
an extra vector $''\, |-1 \rangle\, ''$ $\bigl (|-1\rangle \doteq {\cal F}(K_3) K_- |0\rangle \bigr )$ 
which is annihilated by $K_-$ (it is this requirement that originates the second line in (\ref{Fi})), 
is actually out of ${\mathfrak{F}}$, but out of which $|0\rangle$ can be generated by $K_+$. 
Ket $| -1 \rangle$ has the property that $\langle -1 | -1 \rangle = 1$, and $\langle n\, |\, -1 
\rangle = 0 \; , \; \forall n\geq 0$, leading to the existence of both a $PVM$ and a $POM$ in 
${\cal F}$. The construction of the completed space ${\mathfrak{F}} \cup | -1 \rangle$ can be 
consistently and rigorously carried over resorting to the notion of $''$dilated extension Hilbert 
space$\, ''$ \cite{Sc}. The spectral theory for self-adjoint operators shows indeed that one may 
generally restrict the attention to bounded operators by the use of Cayley transform: if the 
operator is only symmetric, possibly with dense domain, but not self-adjoint, then the Cayley 
transform 
reduces the problem to the analysis of partial isometries. But if the index is non-zero 
({\sl i.e.} the co-dimension of the initial and the final space are unequal) then there will 
not be self-adjoint extensions in the same, given, Hilbert space, but a dilated extension 
Hilbert space is needed for a complete understanding of spectral resolutions and self-adjoint 
operator extensions of unbounded operators. It should be noted that this theory, 
holds for Lie algebras only in the case they are simple, and requires passing 
to the universal enveloping algebra to get the appropriate {$\ast$-representation}. 

From the operational point of view, as in our case co-dimension 1 is sufficient, a 
plausible way to perform the extension is to recall that the principal continuous 
series of class I and the supplementary series have representations spanned by a basis 
$\{ |\lambda , \mu \rangle \}$, corresponding to ${\cal C}_2 = |\lambda |^2 + \frac{1}{4}$,  
for which $\lambda$ is respectively a nonnegative real number and a pure imaginary 
$\lambda = i \tau$, $\frac{1}{2} < \tau < \frac{1}{2}$, such that 
\begin{eqnarray}
K_3 |\lambda ,\mu \rangle &=& \mu \, |\lambda ,\mu \rangle \; , \; \mu = 0, \pm 1, \pm 2, 
\dots \; , \nonumber \\   
K_{\pm} |\lambda ,\mu \rangle &=& \bigl ( \,\pm \left ({\scriptstyle{\frac{1}{2}}} - i 
\lambda \right ) + \mu \, \bigr ) \, |\lambda , \mu \pm 1 \rangle \; . \label{rappr} 
\end{eqnarray} 
Eqs. (\ref{rappr}) imply for $\lambda (\tau )\to 0$, where the two representations are 
isomorphic with each other and with the discrete series (positive $\cup$ negative, through the 
identification $\kappa = \frac{1}{2} + i\tau$), ${\cal C}_2 = \frac{1}{4}$. $|\mu \rangle\equiv 
|0 ,\mu \rangle$ is given, $\forall \mu \geq 1$, by \cite{zero}     
$$ |\mu \rangle = \frac{1}{\mu !}\, K_+^{\mu}\, |0\rangle \; ,\, {\rm but} \; |- \mu 
\rangle = (-)^{\mu} \, \frac{1}{\mu !}\, K_-^{\mu}\, |0\rangle \; . \nonumber$$  
In other words, 
the eigenstates are generated by powers of both $K_+$ and $K_-$. What the desired Hilbert 
space extension amounts to is including in the discrete positive series $|-1\rangle$ 
from one of the two continuous series. 

Physically we suggest the interpretation that whereas $|0\rangle$, even though 
representing a physical $''$vacuum$\, ''$, namely a state with no excitations (zero 
occupation number), yet is still in the space of states of the system, $|-1\rangle$ does 
instead describe the true vacuum, {\sl i.e.} emptyness of the state space of the system, 
out of whose quantum fluctuations such states (in particular $|0\rangle$) can be 
generated.

\end{multicols}

\begin{references}
\bibitem{Di} P.A.M. Dirac, {\sl Proc. Royal Soc. London} A {\bf 114}, 243 (1927) 
\bibitem{SuGl} L. Susskind and J. Glogower, {\sl Physics (N.Y.)} {\bf 1}, 49 (1964) 
\bibitem{CaNi} P. Carruthers, and M.M. Nieto, {\sl Rev. Mod. Phys.} {\bf 40}, 411 
(1968)  
\bibitem{Jac} R. Jackiv, {\sl J. Math. Phys.} {\bf 9}, 339 (1968) 
\bibitem{Ly} R. Lynch, {\sl Physics Reports} {\bf 256}, 367 (1995) 
\bibitem{DuHeSm} D.A. Dubin, M.A. Hennings, and T.B. Smith, {\sl Int. J. Mod. Phys.} 
{\bf 9 B}, 2597 (1995) 
\bibitem{BaPe} V. Barone and V. Penna, {\sl Mod. Phys. Lett.} B {\bf 9}, 685 (1995) 
\bibitem{PeBa} D.T. Pegg, and S.M. Barnett, {\sl Phys. Rev.} A {\bf 39}, 1665 (1989)  
D.T. Pegg, and S.M. Barnett, {\sl J. Mod. Optics} {\bf 44}, 225 (1997)
\bibitem{RoRo} P. Roy and B. Roy, {\sl Quantum Semiclass. Opt.} {\bf 9}, L37 (1997)
\bibitem{WeVoOp} D.-G. Welsh, W. Vogel, and T. Opatrn\'y, {\sl Progr. Optics} {\bf 39}, 
63 (1999) 
\bibitem{Go} G. Gour, {\it The quantum phase problem: steps towards a resolution}, 
{\tt arXiv: quant-ph/0107122 v2}, Mar 2002 
\bibitem{others} It has also been pointed out that the commutation relation $\bigl [ 
{\hat{\phi}},{\hat{n}} \bigr ] =i$ can be implemented as an operator identity only on 
a dense set of the Hilbert space: \\ 
P. Jordan, {\sl Zeits. Physik} {\bf 44}, 1 (1927) \\ 
J.C. Garrison, and J. Wong, {\sl J. Math. Phys.} {\bf 11}, 2242 (1970) \\ 
A. Galindo, {\sl Lett. Math. Phys.} {\bf 8}, 495 (1984) \\
K. Sch\"onhammer, {\it Canonically conjugate pairs and phase operators}, {\tt arXiv: 
quant-ph/0204139}, Apr 2002 \\ 
Such question does not bear on the discussion to follow.    
\bibitem{Pa} M.G.A. Paris, {\sl Fizika} B {\bf 6}, 63 (1997)
\bibitem{Ell} D. Ellinas, {\sl J. Math. Phys.} {\bf 32}, 135 (1991) 
\bibitem{Ban} M. Ban, {\sl J. Math. Phys.} {\bf 32}, 3077 (1991) \\ 
M. Ban, {\sl Opt. Commun.} {\bf 94}, 231 (1992)   
\bibitem{Ka} H.A. Kastrup, {\it How to quantize phases and moduli!}, {\tt arXiv:} 
{\tt quant-ph/0109013}, Sep 2001 
\bibitem{Pe} A. Perelomov, {\it Generalized Coherent States and Their Applications}, 
Springer-Verlag, Basel, 1986 
\bibitem{su(11)}  It is worth recalling that besides the discrete series, one of the 
principal continuous series (class I) and the supplementary series have this property. 
\bibitem{RaTaZe} M. Rasetti, E. Tagliati and R. Zecchina, {\sl Phys. Rev.} A {\bf 55}, 
2594 (1997)
\bibitem{AbSt} M. Abramowitz and I. Stegun, {\it Handbook of Mathematical Functions}, 
Dover Publications, New York, 1972 
\bibitem{note} Over ${\mathfrak{F}}$, $[\![ x ]\!]$ can be realized in terms of multiboson 
operators $A_k$, $A_k^{\dagger}$, ${\hat{N}}_k \doteq A_k^{\dagger} A_k$ \cite{RaDaKaSo}, which 
close an algebra $h(1)$, with $\bigl [A_k, {\hat{n}} \bigr ]= k A_k$, $\bigl [ A_k^{\dagger}, 
{\hat{n}} \bigr ] = -k A_k^{\dagger}$. For $n\equiv uk+v$, where $\displaystyle{u\doteq 
\left [\!\!\left [ \frac{n}{k} \right ]\!\!\right ]}$, and $v\doteq\{ n \}_k$ is the 
residue of $n$ $({\rm mod} k)$, one has ${\hat{N}}_k |n\rangle = u\, |n\rangle$. $\psi ({\hat{n}})$ 
can be constructed resorting to the integral representation $\displaystyle{\psi ({\hat{n}}) = 
\int_0^{\infty} {\rm d}x \frac{{\rm e}^{-x} - {\rm e}^{x{\hat{n}}}}{1-{\rm e}^{-x}}} - \gamma$.  
\bibitem{RaDaKaSo} M. Rasetti, G. D'Ariano, J. Katriel, and A.I. Solomon, 
{\sl J. Opt. Soc. Am.} B{\bf 4}, 1728 (1987)
\bibitem{Sc} K. Schm\"udgen, {\it Unbounded operator algebra and representation theory}, 
Birkh\"auser Akademie Verlag, Basel, 1990  
\bibitem{zero} Uniqueness of $|0\rangle$ imnplies in this representation new relations  
for $K_-$, $K_+$: $2^{\mu} \bigl [(2\mu -1)!! \bigr ]^{-1} \, K_-^{\mu}\, 
|\mu \rangle \equiv |0\rangle \equiv (-2)^{\mu} \bigl [ (2\mu -1)!! \bigr ]^{-1}\, K_+^{\mu} 
\, |-\mu \rangle \; .$
\end{references}
\end{document}